\documentclass[12pt]{article}
\usepackage[a4paper]{geometry}
\usepackage[english]{babel}

\usepackage[pagebackref,draft=false]{hyperref}
\hypersetup{colorlinks,
	linkcolor=myrefcolor,
	citecolor=mycitecolor,
	urlcolor=myurlcolor}

\usepackage{caption}
\usepackage{etaremune}

\usepackage{xcolor}
\definecolor{myurlcolor}{rgb}{0,0,0.4}
\definecolor{mycitecolor}{rgb}{0,0.5,0}
\definecolor{myrefcolor}{rgb}{0.5,0,0}
\usepackage{graphicx}
\usepackage{tikz}
\usepackage{tikz-cd}
\usepackage{mathrsfs}

\usepackage{etoolbox}
\usepackage{makeidx}
\usepackage{sectsty}
\usepackage{dsfont}
\usepackage{enumitem} 
\usepackage[]{latexsym}
\usepackage{braket}
\usepackage{caption}
\usepackage[utf8]{inputenx}
\usepackage[T1]{fontenc}
\usepackage{lmodern}
\usepackage{textcomp}
\usepackage{microtype}
\usepackage{totcount}
\usepackage{blindtext}

\usepackage{fancyhdr}
\usepackage{amsmath}
\usepackage{amsfonts}
\usepackage{amstext}
\usepackage{amssymb}
\usepackage{amscd}

\usepackage{xcolor}
\usepackage{mathrsfs}
\usepackage{tikz}
\usepackage{tikz-cd}

\usepackage{fancyhdr}

\newcommand{\be}{\begin{equation}}
\newcommand{\ee}{\end{equation}}

\newcommand{\bea}{\begin{eqnarray}}
\newcommand{\eea}{\end{eqnarray}}

\newcommand{\blue}[1]{\textcolor{blue}{{#1}}}

\newcommand{\ra}{\rightarrow}

\newcommand{\hh}{\mathcal{H}}
\newcommand{\bh}{\mathcal{B}(\mathcal{H})}

\newcommand{\Slh}{\mathcal{SL}(\mathcal{H})}
\newcommand{\slh}{\mathfrak{sl}(\mathcal{H})}
\newcommand{\Uh}{\mathcal{U}(\mathcal{H})}
\newcommand{\uh}{\mathfrak{u}(\mathcal{H})}

\newcommand{\SUh}{\mathcal{SU}(\mathcal{H})}

\newcommand{\suh}{\mathfrak{su}(\mathcal{H})}

\newcommand{\stsph}{\mathscr{S}(\mathcal{H})}
\newcommand{\Tr}{\textit{Tr}}

\newcommand{\stsp}{\mathscr{S}}

\newcommand{\gr}{\mathrm{g}}

\newcommand{\Gg}{\mathrm{G}}

\begin{document}

	\title{Group actions and monotone metric tensors: \\ The qubit case}

	\author{F. M. Ciaglia$^{1,2}$ \href{https://orcid.org/0000-0002-8987-1181}{\includegraphics[scale=0.7]{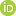}}, F. Di Nocera$^{1,3}$\href{https://orcid.org/0000-0002-1415-2422}{\includegraphics[scale=0.7]{ORCID.png}}, \\
		\footnotesize{$^{1}$\textit{ Max Planck Institute for Mathematics in the Sciences, Leipzig, Germany}} \\
		\footnotesize{$^{2}$\textit{ e-mail: \texttt{florio.m.ciaglia[at]gmail.com} and \texttt{ciaglia[at]mis.mpg.de}}}\\
		\footnotesize{$^{3}$\textit{ e-mail: \texttt{fabiodncr[at]gmail.com} and \texttt{dinocer[at]mis.mpg.de}}} \\ 
	}
	
	\maketitle

\begin{abstract}
In recent works, a link between group actions and information metrics on the space of faithful quantum states has been highlighted in particular cases. In this contribution, we give a complete discussion of this instance for the particular case of the qubit.
\end{abstract}

\section{Introduction}

Because of Wigner's theorem,  it is difficult to overestimate the role of the unitary group $\Uh$ when dealing with symmetries in standard quantum mechanics.
Consequently, it is not surprising that, in the context of quantum information geometry,  $\Uh$ again plays a prominent role when dealing with symmetries.

More specifically, let $\hh$ be the Hilbert space of a finite-level quantum system.
The space of quantum states of this system is denoted by $\overline{\stsp(\hh)}$ and consists of all density operators on $\hh$, that is, $\rho\in\overline{\stsp(\hh)}$ is such that $\rho\geq 0$ and $Tr(\rho)=1$.
Clearly, $\overline{\stsp(\hh)}$  is a convex set, and its interior is denoted by  $\stsp(\hh)$ and consists of density operators on $\hh$ that are invertible, that is, $\rho\in\stsp(\hh)$ is such that $\rho>0$.
It is well-known  that $\stsp(\hh)$ is a smooth manifold whose topological closure is $\overline{\stsp(\hh)}$ \cite{C-I-J-M-2019,G-K-M-2006}, and the tangent space $T_{\rho}\stsp(\hh)$ may be identified with the space of self-adjoint operators on $\hh$ with vanishing trace.
In the context of quantum information geometry, if $\mathcal{K}$ is the Hilbert space of another finite-level quantum system, the allowed quantum channels between the two systems are mathematically described by the so-called completely-positive, trace-preserving (CPTP) maps between $\bh$ and $\mathcal{B}(\mathcal{K})$ sending $\stsp(\hh)$ into $\stsp(\mathcal{K})$  \cite{Holevo-2001}.
These maps are linear on the whole $\bh$, and the standard action of $\Uh$ on $\bh$ given by $\mathbf{a}\mapsto\mathbf{U}\mathbf{a}\mathbf{U}^{\dagger}$ is easily seen to give rise to a CPTP map from $\bh$ into itself for every $\mathbf{U}\in\Uh$.
Moreover, it can be proved that, modulo isomorphisms between Hilbert spaces, a CPTP map $\Phi$ is invertible if and only if $\Phi(\mathbf{a})=\mathbf{U}\mathbf{a}\mathbf{U}^{\dagger}$ for some $\mathbf{U}\in\Uh$.

Following what Cencov did for the classical case of probability distributions and Markov maps \cite{Cencov-1982}, Petz classified the family  of all the Riemannian metric tensors $\Gg^{\hh}$ on every $\stsp(\hh)$ satisfying the monotonicity property
\be\label{eqn: monotonicity property}
\Gg_{\Phi(\rho)}^{\mathcal{K}}\left(T_{\rho}\Phi(\mathbf{a}),T_{\rho}\Phi(\mathbf{a})\right)\,\leq\,\Gg_{\rho}^{\hh}\left(\mathbf{a},\mathbf{a}\right),
\ee
where $\Phi\colon\bh\ra\mathcal{B}(\mathcal{K})$ is a CPTP map sending $\stsp(\hh)$ into $\stsp(\mathcal{K})$, and $\mathbf{a}\in T_{\rho}\stsp(\hh)$ \cite{Petz-1996}.
Unlike the classical case, it turns out that $\Gg^{\hh}$ is not unique, and there is a one-to-one correspondence between the metric tensors satisfying the monotonicity property in equation \eqref{eqn: monotonicity property} and the operator monotone functions $f:\mathbb{R}^{+}\ra\mathbb{R}$ satisfying $f(1)=1$ and $f(t)=t f(t^{-1})$.

Moreover, since $\Phi(\mathbf{a})=\mathbf{U}\mathbf{a}\mathbf{U}^{\dagger}$ is an invertible CPTP map for every $\mathbf{U}\in\Uh$, it follows that every monotone quantum metric tensor must be invariant with respect to the standard action of $\Uh$, and thus the unitary group may be thought of as a universal symmetry group for quantum information geometry.
From the infinitesimal point of view, this means that the fundamental vector fields of the standard action of $\Uh$ on $\stsp(\hh)$ are Killing vector fields for every monotone quantum metric tensor.
Since  we can write $\mathbf{U}=\mathrm{e}^{\frac{1}{2\imath}\mathbf{a}}$ with $\mathbf{a}$ a self-adjoint operator on $\hh$, it follows that the Lie algebra $\uh$ of $\Uh$ may be identified with the space of self-adjoint operators on $\hh$ endowed with the Lie bracket
\be\label{eqn: Lie bracket suh}
[\mathbf{a},\mathbf{b}]\,:=\,\frac{1}{2\imath}\left(\mathbf{ab} - \mathbf{ba}\right),
\ee
and the fundamental vector fields of the action of $\Uh$ on $\stsp(\hh)$ may be labelled by self-adjoint elements in $\bh$.
We will write these vector fields as $X_{\mathbf{b}}^{\hh}$ with $\mathbf{b}\in\uh$.

It is easy to see that the fundamental vector field associated with the identity elements vanishes identically on $\stsp(\hh)$.
Therefore, in the following, we will focus on the special unitary group $\SUh$ rather than on $\Uh$.
This is the Lie subgroup of $\Uh$ generated by the Lie subalgebra $\suh$ consisting of traceless elements in $\uh$.

Elements in  the Lie algebra $\suh$ are not only associated with the fundamental vector fields of $\SUh$ on $\stsp(\hh)$, they are also associated with the  functions $l_{\mathbf{a}}^{\hh}$ on $\stsp(\hh)$ given by
\be
l_{\mathbf{a}}^{\hh}(\rho)\,:=\,\Tr_{\hh}\left(\rho\,\mathbf{a}\right).
\ee
According to the standard postulate of quantum mechanics, every such function $l_{\mathbf{a}}^{\hh}$ provides the expectation value of the (linear) observable represented $\mathbf{a}$, and thus plays the role of a quantum random variable.

It was observed in \cite{C-J-S-2020} that, when the so-called Bures-Helstrom metric tensor $\Gg^{\hh}_{BH}$ is selected among the monotone quantum metric tensors, the gradient vector fields associated with the expectation value functions are complete and, together with the fundamental vector fields of $\SUh$, close on an anti-representation of the Lie algebra $\slh$  integrating to a transitive left action of the special linear group $\Slh$ given by
\be\label{eqn: SLH action BH}
\rho\,\mapsto\,\frac{\gr\,\rho\,\gr^{\dagger}}{\Tr_{\hh}(\gr\rho\gr^{\dagger})},
\ee
where $\gr=\mathrm{e}^{\frac{1}{2}(\mathbf{a} - \imath\mathbf{b})}$, with $\mathbf{a},\mathbf{b}$ self-adjoint operators on $\hh$.

In \cite{Ciaglia-2020}, it was observed that the Bures-Helstrom metric tensor is not the only monotone quantum metric tensor for which a similar instance is verified.
Indeed, if the so-called Wigner-Yanase metric tensor is selected,   the gradient vector fields associated with the expectation value functions are complete and, together with the fundamental vector fields of $\SUh$, close on an anti-representation of the Lie algebra $\slh$  integrating to a transitive left action of the special linear group $\Slh$ which is different than that in equation \eqref{eqn: SLH action BH} and is given by
\be\label{eqn: SLH action WY}
\rho\,\mapsto\,\frac{\left(\gr\,\sqrt{\rho}\,\gr^{\dagger}\right)^{2}}{\Tr_{\hh}\left(\left(\gr\,\sqrt{\rho}\,\gr^{\dagger}\right)^{2}\right)},
\ee
where $\gr=\mathrm{e}^{\frac{1}{2}(\mathbf{a} - \imath\mathbf{b})}$, with $\mathbf{a},\mathbf{b}$ self-adjoint operators on $\hh$.

Moreover, if the so-called Bogoliubov-Kubo-Mori metric tensor is selected, the gradient vector fields associated with the expectation value functions are complete, commute among themselves, and, together with the fundamental vector fields of $\SUh$, close on an anti-representation of the Lie algebra of the cotangent group $T^{*}\SUh$ which integrates to a transitive left action given by
\be\label{BKMaction}
\rho\,\mapsto\,\frac{\mathrm{e}^{\mathbf{U}\ln(\rho)\mathbf{U}^{\dagger} + \mathbf{a}}}{\Tr_{\hh}\left(\mathrm{e}^{\mathbf{U}\ln(\rho)\mathbf{U}^{\dagger} + \mathbf{a}}\right)},
\ee
where $\mathbf{U}=\mathrm{e}^{\frac{1}{2\imath} \mathbf{b}}$, and $\mathbf{a},\mathbf{b}$ are self-adjoint operators on $\hh$.

It is then natural to ask if these three instances are just isolated mathematical coincidences valid only for these very special monotone quantum metric tensors, or if there is an underlying geometric picture waiting to be uncovered and studied.
Specifically, we would like to first classify all those monotone quantum metric tensors for which the gradient vector fields associated with the expectation value functions, together with the fundamental vector fields of $\SUh$, provide a realization of some   Lie algebra integrating to a group action.
Then, we would like to understand the geometric significance (if any)  of the gradient vector fields thus obtained  with respect to the dualistic structures and their geodesics, with respect to  quantum estimation theory, and with respect to  other relevant structures and tasks in quantum information geometry (a first hint for   the Bogoliubov-Kubo-Mori metric tensor may be found in \cite{A-L-2020,F-M-A-2019,Naudts-2021}).
Finally, we would like to understand  the  role (if any) of the ``new'' Lie groups and Lie algebras appearing in quantum information theory because of these constructions.

At the moment, we do not have a satisfying understanding of the global picture, but we will give a complete classification of the group actions associated with monotone metrics for the case of the simplest quantum system, namely, the qubit.

\section{Group actions and monotone metrics for the qubit}

In the case of a qubit, we have $\hh\cong\mathbb{C}^{2}$, and by selecting an orthonormal basis on $\hh$, say $\{|1\rangle, |2\rangle\}$, a basis in $\suh$ is given by the Pauli operators
\be
\sigma_{1}=|1\rangle\langle 2| + |2\rangle\langle 1|,\quad\sigma_{2}=\imath|1\rangle\langle 2|  -\imath|2\rangle\langle 1|,\quad\sigma_{3}=|1\rangle\langle 1| - |2\rangle\langle 2|,
\ee
so that we have
\be\label{eqn: commutators su(2)}
[\sigma_{1},\sigma_{2}]= \sigma_{3},\quad [\sigma_{3},\sigma_{1}]= \sigma_{2},\quad[\sigma_{2},\sigma_{3}]= \sigma_{1}.
\ee
It is well-known that a quantum state $\rho$ may be written as
\be
\rho\,=\,\frac{1}{2}\left(\sigma_{0} + x\sigma_{1} + y\sigma_{2} + z\sigma_{3}\right),
\ee
where $\sigma_{0}$ is the identity operator on $\hh$ and $x,y,z$ satisfy $x^{2} + y^{2} + z^{2}\leq 1$.
If we want $\rho$ to be invertible, then we must impose $x^{2} + y^{2} + z^{2}<1$, which means that $\stsp(\hh)$ is diffeomorphic to the open interior of a 3-ball of radius 1.

Passing from the Cartesian coordinate system $(x,y,z)$ to the spherical coordinate system $(r,\theta,\phi)$, the monotone quantum metric tensor $\Gg_{f}$ associated with the operator monotone function $f$ reads
\be
\Gg_{f}\, = \,\frac{1}{1-r^2}\,dr\otimes dr +\frac{r^2}{1+r}\Biggl(f\biggl(\frac{1-r}{1+r}\biggr)\Biggr)^{-1}(d\theta\otimes d\theta + \sin^2\theta\,\,  d\phi \otimes d\phi) \label{genericg}.
\ee 
In the spherical coordinate system, the fundamental vector fields of $\SUh$ read
\be\label{eqn: Hamiltonian vector fields}
\begin{split}
X_{1} & = -\sin\phi\,\,\partial_\theta - \cot\theta\cos\phi\,\,\partial_\phi \\
X_{2} & = \cos\phi\,\,\partial_\theta - \cot\theta\sin\phi\,\,\partial_\phi \\
X_{3} &=\partial_\phi 
\end{split}
\ee
where we have set $X_{j}\equiv X_{\sigma_{j}}$ for $j=1,2,3$.

For a generic traceless self-adjoint operator $\mathbf{a}=a_{1}\sigma_{1} + a_{2}\sigma_{2} + a_{3}\sigma_{3}$, we have
\be
l_{\boldsymbol{a}}  = \frac{1}{2}(a_1r\sin\theta\cos\phi +a_2r\sin\theta\sin\phi +a_3r\cos\theta)\label{exvalfun},
\ee
and the gradient vector field $Y_{\mathbf{a}}{f}$ associated with $l_{\mathbf{a}}$ by means of $\Gg_{f}$ is 
\be
Y_{\boldsymbol{a}} = G_{f}^{-1} (\mathrm{d}l_{\boldsymbol{a}}, \bullet) = G_{f}^{-1}(\bullet, \mathrm{d}l_{\boldsymbol{a}}).
\ee	
Setting $l_{j}\equiv l_{\sigma_{j}}$ and $Y_{j}^{f}\equiv Y_{\sigma_{j}}^{f}$, straightforward computations bring us to
\be
\begin{split}\label{gradvecfields}
Y_{1}^{f} &=   (1-r^2)\sin\theta\cos\phi\,\,\partial_r   + g(r)\left(\cos\theta\cos\phi\,\,\partial_\theta - \frac{\sin\phi}{\sin\theta}\,\,\partial_\phi\right) \\
Y_{2}^{f} & = (1-r^2)\sin\theta\sin\phi\,\,\partial_r   +g(r)\left(\cos\theta\sin\phi\,\,\partial_\theta +  \frac{\cos\phi}{\sin\theta}\,\,\partial_\phi \right)\\
Y_{3}^{f} &= (1-r^2)\cos\theta\,\,\partial_r - g(r)\sin\theta\,\,\partial_\theta,
\end{split}
\ee
where
\be\label{posg}
g(r) = \frac{1+r}{r}f\biggl(\frac{1-r}{1+r}\biggr) .
\ee
Then, the commutators between the gradient vector fields  can be computed to be:
\be
	\begin{split}\label{fundvecfields}	
		[Y_{1}^{f},Y_{2}^{f}] &= F(r)\,\,X_{3}\\
		[Y_{2}^{f},Y_{3}^{f}] &= F(r)\,\,X_{1} \\
		[Y_{3}^{f},Y_{1}^{f}] &= F(r)\,\,X_{2} 
	\end{split}
\ee
where we have used the expressions in equation \eqref{eqn: Hamiltonian vector fields}, and we have set
\be\label{posF}
F(r) = (1-r^2)g'(r) + g^2(r)
\ee
for notational simplicity.

In order to get a representation of a Lie algebra out of the gradient vector fields and the fundamental vector fields of $\SUh$ we must set $F(r)=A$ with $A$ a constant, obtaining the following ordinary differential equation
\be\label{ode}
(1-r^2)g'(r) + g^2(r) = A
\ee
for the function $g(r)$.
Performing the   change of variable $	t = \frac{1-r}{1+r}$ and separating variables, we obtain the  ordinary differential equation
\be\label{odesepvar}
\frac{g'(t)}{g^2(t) - A} = \frac{1}{2t} .
\ee
This ODE has a different behavior depending on the sign of $A$, thus we will deal separately with the cases where $A$ is zero, positive and negative.

\subsection{First case ($A=0$)}

This case is peculiar, since it is the only case in which the gradient vector fields close a Lie algebra by themselves, albeit a commutative one.
In this case, integration of the ODE in equation \eqref{odesepvar} leads to
\be
g_0(t) = \frac{-2}{\log{t} + c},
\ee
with $c$ being an integration constant arising in the solution of the differential equation.
Taking into account equation \eqref{posg}, we obtain the function
\be
f_0(t) = \frac{t-1}{\log{t} + c}.
\ee
If we want $f_{0}$ to belong to the class of functions appearing in Petz's classification \cite{Petz-1996}, we need to impose $ \lim_{t \to 1} f_0(t) = 1$,
which is easily seen to impose $c = 0$,  and thus we obtain the function 
\be
f_0(t) = \frac{t-1}{\log{t}}.
\ee
This function is operator monotone and  gives rise to the \emph{Bogoliubov-Kubo-Mori} (BKM) metric \cite{Petz-1994}.
Following the results presented in  \cite{Ciaglia-2020}, we conclude that the fundamental vector fields of $\SUh$, together with the gradient vector  fields associated with the expectation value functions by means of the BKM metric tensor,  provide an anti-representation of the Lie algebra of the cotangent group $T^{*}\SUh$ integrating to the action in equation \eqref{BKMaction}.

 It is worth noting that the BKM metric tensor also appears as the Hessian metric obtained from the von Neumann entropy following  \cite{Balian-1999,Balian-2013,Balian-2014}, and this instance has profound implications from the point of view of Quantum Information Theory. 

\subsection{Second case ($A>0$)}

In the case where $A>0$, integration of the ODE in equation \eqref{odesepvar},  together with the  expression of $g(r)$ in equation \eqref{posg}, leads to
\be
f_A(t)=\frac{\sqrt{A}}{2}(1-t)\frac{1+e^{2c\sqrt{A}}t^{\sqrt{A}}}{1-e^{2c'\sqrt{A}}t^{\sqrt{A}}},
\ee
where $c$ is an integration constant arising in the solution of the differential equation.
If we want $f_{A}$ to belong to the class of functions appearing in Petz's classification \cite{Petz-1996}, we need to impose $ \lim_{t \to 1} f_A(t) = 1$, which implies  $c=0$, and thus
\be\label{fA>0}
f_A(t) = \frac{\sqrt{A}}{2}(1-t)\frac{1+t^{\sqrt{A}}}{1-t^{\sqrt{A}}}.
\ee
We now consider the gradient vector fields
\be\label{eqn: gradient vector fields for A>0}
Y_{\mathbf{a}}^{A}\,:=\,G_{f_{A}}^{-1}\left(\frac{1}{\sqrt{A}}\,\mathrm{d}l_{\mathbf{a}}\, , \,\bullet \, \right)
\ee
associated with the (renormalized) expectation value functions $\frac{1}{\sqrt{A}}\, l_{\mathbf{a}}$ by means of the monotone metric tensor $G_{f_{A}}$ with $f_{A}$  given in equation \eqref{fA>0}.
Recalling equation  \eqref{fundvecfields} and the fact that $F(r)=A>0$, we conclude that the fundamental vector fields of the action of $\SUh$ given in equation \eqref{eqn: Hamiltonian vector fields}, together with the gradient vector fields given in equation \eqref{eqn: gradient vector fields for A>0}, provide an anti-representation of the Lie algebra $\slh$ of $\Slh$.
Exploiting  theorem 5.3.1  in \cite{Bhatia-2007}, it is possible to prove that the anti-representation of $\slh$  integrates to an action of $\Slh$ on $\stsph$ given by
\be\label{actionA}
	\alpha_A(\gr, \rho) = \frac{\left(\gr \rho^{\sqrt{A}} \gr^\dagger\right)^{\frac{1}{\sqrt{A}}}}{\Tr\left(\gr \rho^{\sqrt{A}} \gr^\dagger\right)^{\frac{1}{\sqrt{A}}}}
\ee
Notice that when $A=1$, this action reduces exactly to the action of $\Slh$ in equation \eqref{eqn: SLH action BH}.
Accordingly, the associated monotone metric $G_{f_{1}}$  turns out to be the Bures-Helstrom metric, in agreement with \cite{C-J-S-2020}.
A similar discussion can be carried out in the case $A=\frac{1}{4}$. 
In this case, we get the action in equation \eqref{eqn: SLH action WY} and the monotone metric given by this choice is Wigner-Yanase metric.
Again, this is in agreement with results in \cite{Ciaglia-2020}.
The function $f_A$ is continuous on $[0,\infty]$ and differentiable with continuous derivative in $(0,\infty)$ for all $A > 0$. Also, whenever $A > 1$, its derivative tends to the value $-\sqrt{A}/2$ for $t$ approaching 0 from the right. From these two considerations follows that  $f_{A}$ can not be operator monotone whenever $A > 1$, since there exists an interval in which its derivative is negative.
Let us stress that it is still an open problem to find all values $0< A < 1$ for which the function $f_{A}$ is operator monotone. 

\subsection{Third case ($A<0$)}

In this case, setting $B=\frac{-A}{4}$, the integration of the ODE in equation \eqref{odesepvar}, leads us to
\be
	f_{B}(t) = \sqrt{B} \frac{1-t}{2} \tan{\left(\sqrt{B} \log{t} - \sqrt{B} c \right) } ,
\ee
where $c$ is an integration constant.
Since $\stsph$ is diffeomorphic to the open interior of a 3-ball of radius 1, we have that $0 \le r < 1$ and thus $0 < t \le 1$.
Consequently, in this range for $t$, the function $f_{B}(t)$ presents a countable infinity of points where it is not defined, regardless of the values of $B$ and $c$. 
Thus we are led to  exclude this case, which would give rise to a metric tensor $\mathrm{G}_{f_{B}}$ that is not defined everywhere on the space of quantum states $\stsph$.

\section{Conclusions}

Guided by results presented in \cite{Ciaglia-2020} and \cite{C-J-S-2020}, we investigated the relation between the fundamental vector fields of the canonical action of $\SUh$ on the space of faithful quantum states and the gradient vector fields associated with the expectation value functions by means of the monotone metric tensors classified by Petz.
In particular, we give a complete classification of all those metric tensors for which the above-mentioned gradient vector fields, together with the fundamental vector fields of the canonical action of $\SUh$, provide an anti-representation of a Lie algebra having $\suh$ as a Lie subalgebra.
We found that there are only two possible such Lie algebras, namely, the Lie algebra of the cotangent Lie group $T^{*}\SUh$, and the Lie algebra of $\Slh$ (the complexification of $\suh$).

In the first case, there is only one monotone metric tensor associated with the Lie algebra anti-representation, namely the Bogoliubov-Kubo-Mori metric tensor.
Also, this is the only case in which the gradient vector fields close a (commutative) Lie algebra.

In the second case, we found an infinite number of metric tensors associated with an anti-representation of $\slh$.
These metric tensors are labelled by the function $f_{A}$, with $A>0$, given in equation  \ref{fA>0}, and, among them, we recover the Bures-Helstrom metric tensor when $A=1$, and the Wigner-Yanase metric tensor when $A=\frac{1}{4}$, in accordance with the results in  \cite{Ciaglia-2020}.
As already mentioned, we know that $f_{A}$ is never operator monotone when $A > 1$, however, besides the cases $A=1$ and $A=\frac{1}{4}$, it is still an open problem to find all values $0<A<1$ for which the function $f_{A}$ is operator monotone. 
We also found that all these Lie algebra anti-representations integrate to Lie group actions. The explicit forms of these actions are given in equation \eqref{BKMaction} and equation \eqref{actionA}.

The analysis presented here is clearly preliminar because it addresses only the qubit case.
However, building upon the results obtained for the qubit, we have been able to recover the group actions presented here also  in the case  of a quantum system in arbitrary finite dimension.
We are not able to include these  cases in this contribution because of space constraints.
Also,  while for the qubit case we showed that the group actions of $\Slh$ and $T^{*}\SUh$ are the only possible, for an arbitrary finite-dimensional quantum system, we are only able to show that these actions are there, but not that they exhaust all the possibile group actions compatible with monotone quantum metrics in the sense specified before.
This clearly calls for a more deep investigation we plan to pursue in future publications.

%
%
%
\bibliographystyle{splncs04}


\end{document}